\documentclass[usenatbib,onecolumn,12pt]{article}

\usepackage{amsmath,amssymb,graphicx,setspace}

\newcommand{\kepler}{\textit{Kepler}}

\begin{document}

\thispagestyle{empty}


\begin{center}

{\Large The crucial role of ground-based, Doppler measurements for the future of exoplanet science}


\vskip1.0in

Jason H. Steffen$^1$, Peter Plavchan$^2$, Timothy Brown$^3$, Eric B. Ford$^4$,\\
Andrew W. Howard$^5$, Hannah Jang-Condell$^6$, David W. Latham$^7$, Jack J. Lissauer$^8$,\\
Benjamin E. Nelson$^9$, Patrick Newman$^2$, and Darin Ragozzine$^{10}$\\

\bigskip

\footnotesize{
$^{1}$University of Nevada, Las Vegas, Las Vegas, NV 89154\\
$^{2}$George Mason University, Fairfax, VA 22030\\
$^{3}$Las Cumbres Observatory Global Telescope, Goleta, CA 93117, USA\\
$^{4}$The Pennsylvania State University, University Park, PA, 16802\\
$^{5}$California Institute of Technology, Pasadena, CA 91125, USA\\
$^{6}$University of Wyoming, Laramie, WY 82071\\
$^{7}$Harvard-Smithsonian Center for Astrophysics, Cambridge, MA 02138, USA\\
$^{8}$NASA Ames Research Center, Moffett Field, CA 94035\\
$^{9}$Northwestern University, 2145 Sheridan Road, Evanston, IL 60208\\
$^{10}$Brigham Young University, Provo, UT 84602}

\end{center}


\bigskip

\paragraph{Abstract: }  We outline the important role that ground-based, Doppler monitoring of exoplanetary systems will play in advancing our theories of planet formation and dynamical evolution.  A census of planetary systems requires a well designed survey to be executed over the course of a decade or longer.  A coordinated survey to monitor several thousand targets each at $\sim$1000 epochs ($\sim$ 3-5 million new observations) will require roughly 40 dedicated spectrographs.  We advocate for improvements in data management, data sharing, analysis techniques, and software testing, as well as possible changes to the funding structures for exoplanet science.




\newpage
\setcounter{page}{1}

\noindent
\textit{I have heard it said that ``the finder of a new elementary particle used to be rewarded by a Nobel Prize, but such a discovery now ought to be punished by a \$10,000 fine''.}

\hskip1.0in --- Nobel Laureate Willis E. Lamb, Jr.
\vskip0.1in


\paragraph{Advancing theory with whole system architectures} \mbox{}

A general theory of planet formation and dynamical evolution requires a clear understanding of the statistical properties of the global population of planetary systems.  Ultimately, we need a theory that goes beyond just reproducing existing observations, but one that is also able to predict the results of observations we might endeavor to make---such as the number of potentially Earth-like planets with an outer system of gas giants in the solar neighborhood.  There is currently a significant, world-wide effort to measure and characterize the physical properties of individual exoplanets---particularly those that are small.  And, while the study of individual planets is an important endeavor, their characterization is only a portion of the scientific value that exoplanet discoveries present.

Some of the key observables of the process of planet formation, and a system's dynamical evolution, are the types of planets and systems that form through those processes.  Successfully reconstructing the evolutionary paths that systems may take tells us of the physics that occurs at various points along the way.  Ultimately, the variety of initial conditions in the disk and the environment must produce the variety of observed systems with the correct relative frequencies---the ``branching ratios'' of the different outcomes.  To fully understand the nature and history of planetary systems, we must establish the variety of architectures of those systems.  That is, we must determine the existence of different populations of planetary systems, know how those systems are characterized, and understand how the planets in those systems correlate with each other and with the host star.  Stellar Radial Velocity (RV), or Doppler, measurements are essential for achieving these goals.

Knowing how planets in a system are arranged, how that arrangement depends upon stellar or environmental properties, how system architecture changes with distance from the host star, and how the inner planets are related to the outer planets, are just a few examples of information that will have a material impact on our theories.  Establishing the different system architectures should be a high priority in the coming years and a robust ground-based Doppler campaign, with a similarly robust analysis of its results, is crucial for building a general, predictive planet formation theory---especially given the limitations of the various alternative planet detection methods.

For RV measurements to fill their scientific potential, there must be homogeneous observational programs that focus on the demographics of planetary systems.  If RV measurements are allocated simply to chase the lowest mass or most-Earth-like-planet to date, the whole endeavor runs the risk of losing its scientific direction---essentially becoming a tool with limited scope rather than a fully developed program in its own right.  Moreover, such \textit{ad hoc} programs for individual systems render the observations practically useless for the statistical analyses of populations.

\paragraph{Doppler measurements in the TESS era}\mbox{}

Doppler measurements will play the essential role in measuring the whole-system architectures of planetary systems.  The TESS mission is expected to identify a large sample of $\sim 10,000$ planetary systems \cite{Sullivan:2015}---constituting a large discovery space for future observations since they will be relatively close and bright.  Nevertheless, TESS alone cannot connect the short-period, \kepler-like planets that will be the bulk of its discoveries to planets that lie beyond a few AU.  At the same time, GAIA should discover an even greater number of giant planets, with a peak sensitivity near a few AU \cite{Gaia:2016}.  But, GAIA cannot connect the more distant giant planets to the inner parts of the systems.  Moreover, the most favorable systems with either TESS or GAIA will have several tens of transit or position measurements---not the several hundreds needed to characterize the systems.  Microlensing surveys provide quantities in different mass and separation regimes than existing studies but lack the multi-epoch observations needed to characterize a whole system \cite{Gaudi:2012}.  RV measurements are unique in their ability to fill multiple gaps among the different detection methods and serve as the glue that binds complementary observations together.

To identify and characterize the different types of planetary systems, the scope of the necessary work is staggering.  Figure 1 shows an estimate of the number of planetary systems that a survey must have in order to identify (with 95\% confidence) a second population with a distinguishing statistical feature constituting a given relative fraction and separated from the main population by some distance.  It also shows the duration that a survey must have for a sample of TESS targets given the single-measurement noise.

If TESS finds 10,000 targets (even if only half will be amenable to RV follow-up), then to measure the architectures of these systems---from the multiple, small planets (transiting and non-transiting) near the star out to the decade-long orbits of Jupiter analogs---we would need $\sim 500-1000$ RV measurements per target, or three to five million RV measurements gathered over the next decade.  Assuming 3,000,000 observations, 10 years, 330 nights per year, and 20 measurements per night, then more than 40 dedicated, moderately high precision ($\sim$ m/s) spectrographs are required---just for the science observations.  Making use of those observations requires continued improvements to our understanding of the effects of stellar (and other) noise.

\begin{figure}
\begin{center}
\includegraphics[width=0.3\textwidth]{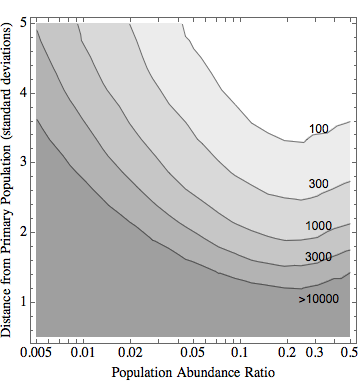} \qquad \qquad
\includegraphics[width=0.45\textwidth]{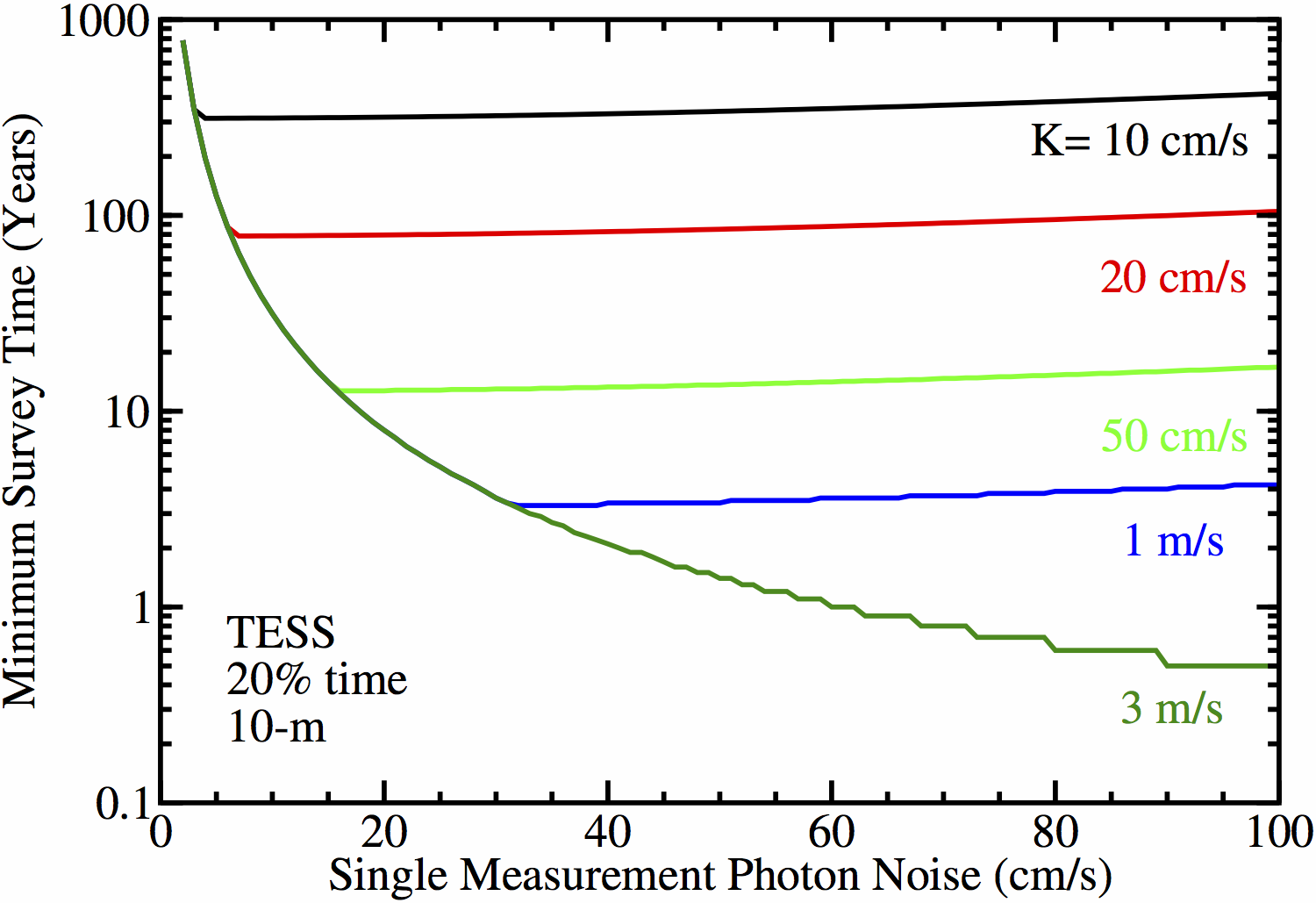}
\caption{Left: Approximate sample size needed to distinguish two normally distributed populations with 95\% confidence as a function of the population separation and relative abundance.  Right: The minimum survey duration as a function of the single measurement precision for velocity semi-amplitudes K for a single planet detection of arbitrary period and SNR=10 for a sample of 62 TESS candidates (V=8--12 mag) assuming $K\propto1/\sqrt{N_{obs}}$., a minimum integration time of 5 min to average p-mode oscillations, a 3-min slew time, a minimum of 10 observations per target, 8 hours of operation per night, and 50\% loss from weather or technical issues.  Actual survey times may be longer due to disentangling stellar activity and multiple planet signals, cadence aliasing, candidate follow-up, etc.\label{RVsim}}
\end{center}
\end{figure}

The effort to calibrate and reduce the data, analyze them to measure or constrain additional planets, to correct for biases, determine the distributions of orbital properties, to identify correlations among the various parameters, and to interpret those results is an effort comparable in scope to any other major aspect of the survey.  Such a program requires a significant degree of industrialization be incorporated into to the process.  A global coordination of observations, the division and specialization of labor, unbiased and timely reporting of results and analysis, large-scale calibration of both hardware and software, and a transparent means to identify and address needs should be central to the program.

\paragraph{Dissemination and curation of RV data}\mbox{}

As the field of exoplanets moves from the era of discovery where each new planet opens new territory to explore, to the era of understanding where that territory is mapped and its theoretical implications are identified, the demands on the data change.  Without common standards for reporting observational results, the analysis and interpretation of those results is at least challenging if not impossible.

Having a standard for reporting and ingesting RV and stellar data into online archives will be crucial for conducting these population-level analyses.  For example, a recent straightforward analysis of $\sim 100$ systems required several days worth of effort just collecting the correct stellar parameters.  This work was in addition to the process of discovering the personality of the individual data files---which have been tabulated by hand, scouring through the literature.  Currently, the RV data stored in online archives are heterogeneous in their content and format, many systems are missing important stellar parameters, and no information exists regarding the analysis that produced those data (e.g., parameter priors from a Markov Chain or the analysis software used to determine the values and uncertainties).

Raw spectroscopic data should be archived to allow the use of various methods to eliminate stellar activity, to facilitate combining data from different instruments, and to compare the results of different RV measurement techniques.  Metadata about each observation---why and how the decision was made to observe this target at this time---is also missing, but is important for the proper interpretation of the results (e.g., so effects caused by changes to the observing strategy for can be addressed).  After nearly three decades of investment to produce the data that we have, it is unfortunate that the legacy of that work is in such disarray.  We recommend that the exoplanet community devise a standard for reporting raw and reduced data, as well as important metadata, that can be met in an automated fashion by observers, in order to establish a data archive worthy of the science.

Historically there has been significant emphasis on planet discoveries with the effect that non-detections are rarely published.  There may have been some justification for this practice in the past, however as the field matures it becomes counter productive since the practice stands in the way of gaining the deep insights that can only be found through a careful analysis of large, homogeneous samples.  By neglecting the value of null results we lose the ability to understand the fundamental properties of the exoplanets and their systems.  As we aim to measure the architectures of whole planetary systems, we need survey results that are not intentionally biased in unknown ways.

The benefit of producing complete and transparent data products is that such catalogs are the basis for high impact literature.  Among the most highly cited astronomy papers in recent decades are papers describing and characterizing the broad release of scientific data---often without an in-depth analysis for astrophysical signals---such as data releases from the Sloan Digital Sky Survey, WMAP, and \kepler (which often have many thousands of citations).  This same model could be applied to data taken from one or a collection of ground-based spectrographs.  Papers describing the instrument, observations, and providing useful data for analysis are likely to be more influential than a series of smaller papers describing individual systems, and would be far superior to having the data rest dormant on a computer until ``someone'' has the opportunity to analyze the data (while simultaneously biasing the exoplanet sample and risking the data languishing indefinitely---wasting both the effort and the facilities).

\paragraph{Improving analysis methods}\mbox{}

The software used to analyze a large sample of data differs from software that is applied to individual systems---especially as it relates to automation.  Analysis pipelines require significant testing to both characterize and optimize their performance.  Data challenges are a common practice in other disciplines to perform these tests, but have only recently been applied to exoplanet data (e.g., \cite{Dumusque:2017,Street:2018}, and the EPRV3 Evidence Challenge).  Data challenges are most effective when they address specific questions, so that teams can learn the benefits and weaknesses of different methods and assumptions.  A series of well-constructed challenges is valuable primarily for helping to improve the state-of-the-art in the field, rather than picking what is best of the current submissions (i.e., the ``winner'' of the challenge).  Improvement in any technical arena requires considerable effort, both of people designing the challenges and analyzing the results and of the participants.  The current approach to funding makes it difficult to obtain concurrent support for multiple teams from multiple countries.

As the analysis software becomes more complex, it will require greater input from a larger number of individuals.  One or a few standard codes that are open and modular and that are tested via appropriate data challenges is a far better option than having students reinvent RV analysis codes every few years.  These codes provide a better standard with tested and understood capabilities and limitations.  Appropriate modularity allows more scientists to contribute to the development in areas where their expertise is particularly strong.  The primary routines for such codes should be written in an open programming language that is compiled and strongly-typed (e.g., C/C++, Julia).  Wrapper interfaces to those routines from high-level languages (e.g., Python, Julia, or R) can further facilitate broad participation---including researchers from complementary fields such as statistics and computer science.  An example worth considering is the development of the MESA stellar evolution code \cite{Paxton:2011}.

\paragraph{Effective funding structures}\mbox{}

The effort to conduct a broad census of planetary system architectures may require changes to funding structures.  Yet, such a census would significantly enhance the value of space missions and would complement previous work done from ground-based facilities.  To fund a complete survey, including the construction and operation of the essential equipment, and the analysis of the resulting data would be about an order of magnitude larger than current endeavors.  This program would require the consolidation of existing collaborations under a single, organizing umbrella.  Similar consolidation occurred in dark matter detection experiments over the last decade where different technologies were tested at the small and medium scales and the way was prepared for the current, large international programs (with combined budgets nearing \$100 million).

Conceivably, multiple aspects of the proposed survey could be partitioned off into smaller programs.  However, essential elements (such as the development and characterization of an analysis pipeline) are likely to be more expensive than the typical \$300k single-PI grants, but far less expensive than the \$10 million instruments or \$30 million balloon missions (the next largest step).  Programs spanning several professors with their research groups, (e.g., data challenges or stellar systematic characterization) could require multi-year grants with annual budgets on the order of \$1 million.  And, while such programs are likely to be required for the future of ground-based exoplanet science, there are few if any funding structures that meet these needs at an appropriate scale.

\paragraph{Conclusions}\mbox{}

A comprehensive understanding of the properties of whole planetary systems is essential for the development of a complete theory of planet formation.  We recommend that a large-scale, ground-based Doppler program, capable of measuring and quantifying the variety of planetary system architectures out to decade-long orbits, be designed and implemented.  The effective execution of such a survey requires improvements to several aspects of the process including: spectrograph Doppler precision; characterization of stellar noise; survey design and execution; data standards, management, and accessibility; and statistical analysis methods and automation.  The effective mitigation of each of these items may require changes to the current funding structures at both NASA and the NSF so that teams of scientists of adequate size can efficiently work together to resolve the various issues.

\bibliographystyle{unsrt}
{\scriptsize
\begin{spacing}{0.25}
\bibliography{whitebib}
\end{spacing}
}

\end{document}